\documentclass[article,aps,nofootinbib, twocolumn]{revtex4}
\usepackage{graphicx}
\usepackage{amsmath,amssymb}
\usepackage{hyperref}
\usepackage[usenames]{color}

\hypersetup{
   colorlinks=true,
   linkcolor=red,
   citecolor=blue,
}

\def\be{\begin{equation}}
\def\ee{\end{equation}}
\def\ba{\begin{eqnarray}}
\def\ea{\end{eqnarray}}

\begin{document}

\title{Detecting Features in the Dark Energy Equation of State: A Wavelet Approach}

\author{Alireza Hojjati$^{1}$,  Levon Pogosian$^{1}$, Gong-Bo Zhao$^{1,2}$}

\affiliation{$^1$Department of Physics, Simon Fraser University, Burnaby, BC, V5A 1S6, Canada \\
$^2$Institute of Cosmology \& Gravitation, University of Portsmouth, Portsmouth, PO1 3FX, United Kingdom}

\begin{abstract}
We study the utility of wavelets for detecting the redshift
evolution of the dark energy equation of state $w(z)$ from the
combination of supernovae (SNe), CMB and BAO data. We show that local
features in $w$, such as bumps, can be detected efficiently using wavelets. To demonstrate, we
first generate a mock supernovae data sample for a SNAP-like
survey with a bump feature in $w(z)$ hidden in, then successfully
discover it by performing a blind wavelet analysis.
We also apply our method to analyze the recently released
``Constitution'' SNe data, combined with WMAP and BAO from SDSS, and
find weak hints of dark energy dynamics. Namely, we find that models
with $w(z) < -1$ for $0.2 < z < 0.5$, and $w(z)> -1$ for $0.5 < z
<1$, are mildly favored at 95\% confidence level. This is in good
agreement with several recent studies using other methods, such as
redshift binning with principal component analysis (PCA)~(e.g. Zhao
and Zhang, arXiv: 0908.1568).
\end{abstract}

\date{\today}

\maketitle

\section{Introduction}

The observed acceleration of cosmic expansion
\cite{Riess:1998cb,Perlmutter:1998np} means that either our
understanding of gravity on cosmological scales is incomplete, or
that there exists a yet unknown dark energy (DE) component, which is
gravitationally repulsive, contributing around $70$\% of the total
energy budget of the universe. The redshift evolution of the DE
fraction is determined by its equation of state $w(z)$ via \be
\Omega_{\rm DE}(z)= \Omega_{\rm DE} \exp\big\{-3\int d{\rm
ln}z'[1+w(z')]\big\} \ , \label{friedmann} \ee where $\Omega_{\rm
DE}$ is the dark energy fraction at $z=0$. The simplest DE candidate
is the cosmological constant $\Lambda$ with $w=-1$. However, in the
absence of a convincing theory of DE, one should not disregard the
possibility of a more complicated evolution history of $w(z)$.

Much effort has been put in extracting $w(z)$ from data. Even though $\Lambda$ provides a good fit \cite{wmap5-komatsu}, current experimental constraints on $w$ are not accurate enough to rule out the possibility of a time-varying $w$. It is of
fundamental importance to know if DE is a constant or dynamical, and future observations will be in a much better
position to address this important question \cite{albrecht_detf}.
Approaches used so far to estimate $w(z)$ from data can be classified into three types: fitting a model \cite{Linder:2002et,Chevallier:2000qy,corasaniti,alam}, binning and principal component analysis (PCA) \cite{pca-huterer,pca-pogosian,pca-cooray}, and a direct differentiation of distance data  \cite{Starobinsky}. In the first method, one assumes a functional form for $w(z)$ or the Hubble parameter $H(z)$ and fits the form to data. Many different functions have been considered and the results are generally strongly dependent on the assumed model. In the PCA method, recently adopted by the JDEM Figure of Merit Science Working Group \cite{jdem_fom}, $w(z)$ is binned in redshift or scale-factor, and the value in each bin is considered as a parameter of the model. The best measured principal components can be used to reconstruct $w(z)$. In the third method, the Type Ia SNe luminosity distance data is smoothed, and $w(z)$ is calculated from the second derivative of this smoothed function.

All of the three techniques above have their merits, and can be used
to address appropriately formulated questions. In this paper we are
examining the possibility of using wavelets as a tool to detect
possible features in $w(z)$ from a combination of supernovae (SNe),
cosmic microwave background (CMB) and baryon acoustic oscillations
(BAO) data, in a model-independent way.  Local featues in $w(z)$ would exist if, for example, dark energy was a quintessence scalar field with a potential that had a a step-like bump. Sharp features in $w(z)$ are also predicted in certain types of phantom models of dark energy \cite{Carroll:2003st}.

An advantage of wavelets is that one does not need to assume a
prior knowledge of the position and size of features in $w(z)$. Wavelets search for large
and small scale features at the same time and can reconstruct them with only a few parameters.

This paper is organized as follows. In Sec.~\ref{sec:method} we
describe the wavelet method. In Sec.~\ref{sec:future} we demonstrate
the method by hiding a feature in a simulated data and then
performing a blind analysis to reconstruct it. In
Sec.~\ref{sec:current} we apply the same method to search for
features in the current data and conclude with a summary in Sec.~\ref{sec:summary}.

\section{Methodology}

\label{sec:method}

\subsection{The wavelet parametrization}

Wavelets, known as ``mathematical microscopes'', have the ability to
``see'' signals on multiple scales, making them a powerful tool for
detecting features in noisy data. Examples of their use in cosmology
include reconstructing the primordial power spectrum
\cite{Mukherjee, Shafieloo2}, probing the non-Gaussianity of the
primordial fluctuations~\cite{wavelet_nonG}, and searching for
signals of parity breaking~\cite{Cabella:2007br}. Wavelets are
specially designed functions that have the property of being
localized in both frequency and configuration space. Individual
wavelets are generated from a ``mother'' function as \be \psi_{nm}
(x) = \Big(\frac{2^n}{L}\Big) ~ \psi\Big(\frac{2^n~x}{L}-m\Big)
\label{eq:wavelet} \ee where $\psi(x)$ is the mother function, $L$
is the range on which the function is defined,  $0<m<2^n-1$ and
$0<n<N-1$. Each wavelet function has a localized shape. The index
$n$ determines the size of the feature, while $m$
determines its location on $x$. Functions $\psi_{nm}$ are orthogonal
by design, hence we can expand any function onto the wavelet basis
in the same way as we do for any other orthogonal basis:
\be
f(x)=\sum_{n=0}^{N-1} \sum_{m=0}^{2^n-1}  a_{nm} \psi_{nm} (x) \ .
\ee
By this definition, each wavelet acts as a band-pass filter and an
infinite number of wavelets are needed to cover the whole spectrum.
This problem is solved by introducing a scaling function. Wavelets
coefficients together with the coefficient of the scaling function,
$2^N$ in total, can be used to describe $f(x)$ at $2^N$ points.

\begin{figure}[thp]
\includegraphics[scale=0.9]{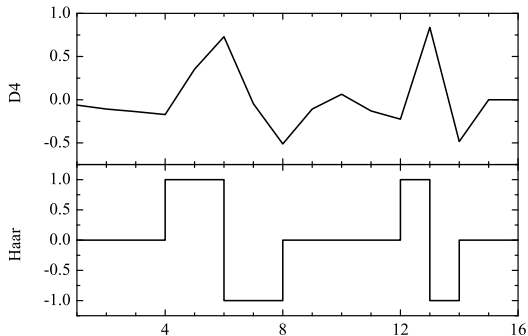}
\caption{The $5$th and $15$th D4 (upper) and Haar (lower) wavelets.
The resolution ($16$ points) in this illustration is the same as the one used in
the numerical calculations in this paper.} \label{fig1}
\end{figure}
In this work, we consider $w(z)$ to be an unknown function which
can, in principle, have local features in $z$. Like any other
function, we can expand it into the wavelet basis as
\be
1+ w (z_j)
=  \sum  P_i \psi_i (z_j) \ ,
\label{w_expansion}
\ee
where $z_j$ are the redshift points at which we calculate $w$,
$P_i$ are the wavelet expansion coefficients, and $\psi_i$ are
the wavelet functions. We chose to simplify our indices so that $P_1$
is the scaling function coefficient, $P_2 = a_{00}$, $P_3= a_{10}$,
etc. We then treat the wavelet expansion coefficients as the
parameters of the model to be measured from data. Note that we
expand $1+ w(z)$, as opposed to $w(z)$, so that all coefficients are
zero if $w=-1$, as in the currently favored $\Lambda$CDM model. We
use Daubechies $4$ (D4) \cite{Daubechies} and Haar \cite{NR}
wavelets (prototypes shown in Fig.~\ref{fig1}) in this paper.  The
D4 wavelets are highly localized and suitable for reconstructing
sharp features. The Haar wavelets are made of step functions. We
should keep in mind that our results using different types of
wavelets should converge and be consistent with each other in the
limit of large number of data points.

Based on the expected redshift range of a future Type Ia SNe survey like
SNAP, we take the redshift range to be $0 < z < 1.7$.  We choose to
work with up to $16$ wavelet parameters, which allows us to calculate
 $w(z)$ at $16$ equally spaced points in $z$ and probe features of up to
 $1.7/8 \approx 0.2$ size in redshift. The values of the wavelet coefficients are determined by fitting
(\ref{w_expansion}) to data using Markov-chain Monte Carlo (MCMC) as
explained in Subsection \ref{sec:optimized}.

\begin{table*}[tbh]
\begin{center}
\begin{tabular}{|@{~}c@{~}|l@{     }|c@{   }c@{   }c@{   }c@{   }c@{   }c@{   }c@{   }c@{  }
c@{   }c@{   }c@{   }c@{   }c@{   }c@{   }c@{   }c@{   }c|@{}}
\hline \multicolumn{1}{|c}{}&\multicolumn{1}{r|}{$z \rightarrow$}
&~0.1~&~0.2~&~0.3~&~0.4~&~0.5~
&~0.6~&~0.7~&~0.8~&~0.9~&~1.0~&~1.1~&~1.2~&~1.3~&~1.4~&~1.5~&~1.6~&~1.7\\
\hline JDEM & $N(z)$
&~300~&~35~&~64~&~95~&~124~&~150~&~171~&~183~&~179~&~170~&~155~&~142~&~130~&~119~
&~107~&~94~&~80\\
& $\sigma_m(z)$ ($\times
10^{-3}$)~&~9~&~25~&~19~&~16~&~14~&~14~&~14~&~14~&~15~&~16~&~17~
&~18~&~20~&~21~&~22~&~23~&~26\\
\hline
\end{tabular}
\end{center}
\caption{The redshift distribution of type Ia supernovae $N(z)$ for
a SNAP-like JDEM survey \cite{kim_etal}, together with 300 SNe from the NSNF \cite{nsnf}. The
redshifts given are the upper limits of each bin. Magnitude errors
$\sigma_m(z)$ are evaluated at bin midpoints.}
\label{tab:snap}
\end{table*}
\subsection{How many wavelet coefficients?}
\label{sec:optimized}

The main advantage of wavelets is that we do not need to a
prior knowledge of the position, the scale and the shape of the
possible features in the function $w(z)$. Wavelets search for large
and small scale features at the same time and the fact that they are
localized in configuration space makes it possible to reconstruct
features with only a few nonzero coefficients. In practice, we
start with performing an initial MCMC run with all $16$ wavelet
coefficients as free parameters. We then keep those parameters whose
signal-to-noise ratio (SNR) is above a certain threshold, and set
parameters with low SNR to zero, since they are not sensitive to features in $w(z)$.
From the surviving set of wavelet coefficients, we find the combination that has the lowest reduced $\chi^2$.
This gives us the optimal number of
wavelets we need to detect a feature in $w(z)$. We then use three model selection criteria, described below, to determine if this wavelet fit is in fact preferred over a constant $w$.

To judge whether a model is preferred by data, we use three
selection criteria: the improvement in $\chi^2$ per extra degree of
freedom $\Delta \chi^2/\Delta({\rm DoF})$,  the Akaike Information
Criterion (AIC) \cite{AIC}, and the Bayesian Information Criterion
(BIC) \cite{BIC}. The AIC and BIC are defined as \ba
{\rm AIC} &=& -2 \log{\mathcal{L}} + 2 N_p + \frac{2 N_p (N_p +1)}{N_d - N_p -1 }    \\
{\rm BIC} &=& -2 \log{\mathcal{L}} + \frac{N_pN_d
\log(N_d)}{N_d-N_p-1} \ , \ea where $N_p$ is the number of
parameters and $N_d$ is the number of data points. The second terms
on the right hand sides of these relations show the penalty for
adding new parameters, which is stronger for BIC. Both expressions
include second-order corrections, which is important for the case
with a small data sample size, i.e. small $N_d$. The AIC and BIC are
widely used in the model selection -- the lower the value of
AIC(BIC), the more the model is preferred. A model with $\Delta
{\rm BIC} < -10$ is decisively preferred, while models with
$-10\leq\Delta {\rm BIC}<-6$, $-6\leq\Delta {\rm BIC}<-2$ and
$\Delta {\rm BIC}\geq2$ have very strong, strong, positive and weak
evidence of model preference, respectively. In the case of the AIC, models with $\Delta {\rm
AIC} \leq 2$ have substantial support, models with $4 \leq \Delta
{\rm AIC} \leq 7$ have considerably less support and those with
$\Delta {\rm AIC} > 10$ have essentially no support.

\subsection{Current and future data sets}
\label{sec:data}

For our tests we will work with a combination of SNe, CMB and BAO
data, considering both the current available data and the simulated
future data. The combined likelihood of the model is calculated as
\be -2 \log{\mathcal{L}} = \chi^2_{total} = {\chi}^2_{\rm SNe} +
\chi_{\rm CMB}^2 +\chi^2_{\rm BAO} \ee with \be \chi^2 = (\mathbf{x}
- \mathbf{d})^{T} \mathbf{C}^{-1} (\mathbf{x} - \mathbf{d}) \ee
where $\mathbf{x}$ is theoretical value for an observable,
$\mathbf{d}$ is data vector and $\mathbf{C}$ is the data covariance
matrix.

\subsubsection{SNe data}

The observable quantity in SNe data is the redshift-dependent
magnitude: \be m = M + 5 \log{d_L} + 25 \label{def:mz} \ee where $M$
is the intrinsic supernova magnitude and $d_L$ the luminosity
distance. The luminosity distance is defined as \be \label{lumdis}
d_L(z)=(1+z) \int_0^z {d{z}' \over H({z}')} \ , \ee where $H(z)$ is
the Hubble parameter with a current value of $H_0$. We use the
characteristics of a SNAP-like JDEM survey \cite{kim_etal} to simulate futuristic SNe
data with errors given by
\be
\sigma_m(z) =
\sqrt{\frac{\sigma_\mathrm{obs}^2}{N_{\mathrm{bin}}}+ \mathrm d
m^2} \ ,
\label{sigma_m}
\ee
where $\sigma_\mathrm{obs} = 0.15$. The
systematic error, $\mathrm d m$, is assumed to increase linearly
with redshift:
\be
\mathrm d m = \delta m
\frac{z}{z_{\mathrm{max}}},
\ee
$\delta m$ being the expected
uncertainty and $z_{\mathrm{max}}$ the maximum redshift. Table
\ref{tab:snap} shows the redshift distribution of type Ia supernovae
expected to be detected by JDEM together with estimated error in
magnitude for different redshift  bins. In addition, we assume that $300$ supernovae at $z<0.1$ will be
measured by the Nearby Supernovae Factory (NSNF) \cite{nsnf}. For current SNe data, we use
the recently released ``Constitution" sample with 397 data points \cite{Hicken}.

\subsubsection{CMB and BAO data}

Rather than working with the full CMB data, we use the procedure
suggested in \cite{wmap5-komatsu} and work with two distance ratios:
the shift parameter $R$, and the acoustic scale $\ell_A$, defined as
\ba
R(z_*) &=& \frac{\sqrt{\Omega_m H_0^2}}{c} (1+z_*) D_A(z_*) \\
\ell_A &\equiv& (1+z_*) \frac{ \pi D_A(z_*)}{r_s (z_*)} \ea where $z_*$ is the redshift of decoupling, $D_A(z)$
is the proper angular diameter distance to redshift $z$, and $r_s
(z)$ is the comoving sound horizon at $z$. It was shown that these
two quantities contain almost all of the information required to
constrain dark energy with CMB \cite{wang,Elgaroy,Li:2008cj}. For
future CMB data, we used the Planck characteristics for parameters
and their errors \cite{planck}. For today's data we used the WMAP-5
values and errors for the parameters \cite{wmap5-komatsu}.

We use the BAO data point at $z = 0.35$ from \cite{Percival}
corresponding to $r_s/D_V(z)$, where $D_V (z)$ is given by
\be
D_V =
\left[(1+z)^2 D_A^2(z) \frac{c z}{H(z)} \right]^{1/3} \ ,
\ee
where $c$ is the
speed of light.

\subsection{PCA analysis}
\label{sec:pca}

A significant departure from zero of
any of the wavelet coefficients would indicate a departure from the
$\Lambda$CDM model. Although the wavelet functions are orthogonal by
construction, and hence their expansion coefficients should be
uncorrelated, the error bars on the fitted wavelet coefficients are
actually correlated. In the presence of correlation among
parameters, one can not determine conclusively at what significance
level a deviation from zero has occured for any one of the
coefficients. In order to quantify the detection of a departure from
zero we decorrelate the wavelet coefficients. The PCA
analysis \cite{PCA} is designed to do precisely that by
diagonalizing the covariance matrix of the original parameters to
find their uncorrelated linear combinations. The original data
vector $P_j$ is transformed into a new data vector $q_i$ using the
same rotation matrix $\omega_{ij}$ that diagonalizes the covariance
matrix. Namely, we have
\be
q_i = \sum_j \omega_{ij}~ P_j
\ee
so that the new covariance matrix, $\mathbf{C_q} = \mathbf{\omega}~
\mathbf{C_P} ~\mathbf{\omega^{-1}}$, is diagonal, and $q_i$'s are the
uncorrelated principal components of $\mathbf{P}$. Like the wavelet
coefficients, principal components should also be zero for the
$\Lambda$CDM model. We will use the PCA to quantify the significance
with which the current data sees departures from $\Lambda$CDM.

\subsection{Features in $w(z)$}

Broad features in $w(z)$, such as a single transition from one constant
value to another, can be efficiently detected by other methods
such as binning. Our aim is to develop a method to detect, if any, local features in w(z) in a model-independent way. 
In this sense, we found that wavelets can do a good job phenomenologically. 
Wavelet method is more effective for searching for localized features in $w(z)$ that would likely require
a larger number of parameters to describe with other methods.  From the theoretical point of view, a local feature in $w(z)$ is not particularly motivated, but can be produced in scalar field models of dark energy with a step-like bump in the potential, or in certain types of phantom models \cite{Carroll:2003st}.

In the next section we will try to recover a feature in $w(z)$, a
bump, that we ``hide'' in a simulated future data. Such a bump is
not ruled out by current data~\cite{bump}, since the averaged $w(z)$
-- the quantity that the current data essentially constrains -- does
not significantly deviate from $-1$ for this feature. The main rationale for
considering this particular choice is that such a bump feature
cannot be easily discovered by binning if one does not know the
size, the position and the scale of the bump in advance.

\begin{figure}[tbh]
\includegraphics[scale=0.7]{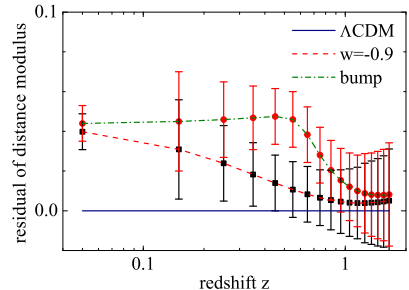}
\caption{Comparison of the residuals of the distance modulus for $\Lambda$CDM, WCDM and bump models with characteristic error-bars for SNAP.}
\label{fig0}
\end{figure}

\section{Reconstructing features with future data}
\label{sec:future}

\begin{figure*}[tbh!]
\includegraphics[scale=0.6]{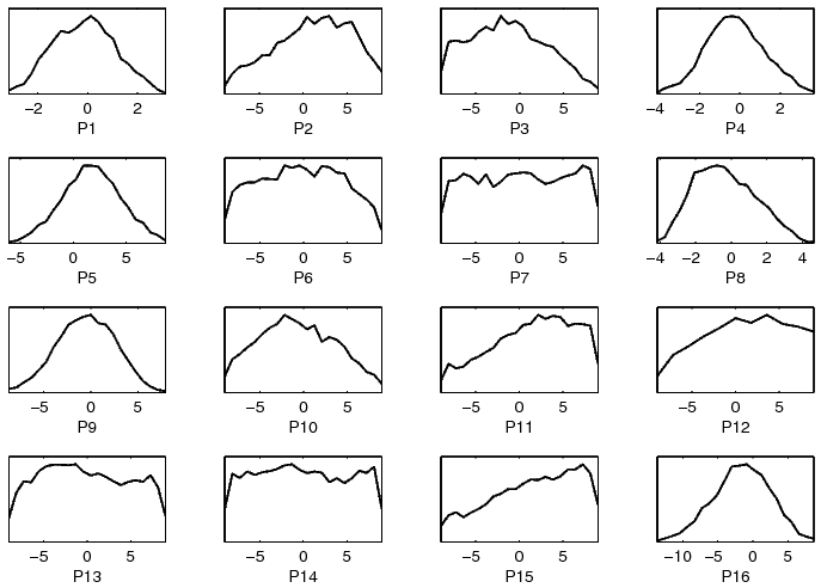}
\caption{The $1$-D posterior distributions of all the $16$ wavelet
coefficients obtained from our first MCMC run.}
\label{fig2}
\end{figure*}

In this section, we use the wavelet method to reconstruct a feature hidden in the simulated mock future data. Our procedure can be summarized as follows:
\begin{enumerate}
\item Generate a mock dataset with a feature of $w(z)$ hidden in;
\item Perform the first MCMC fit for $16$ wavelet coefficients;
\item Sort the wavelet coefficients according to their signal-to-noise ratio (SNR) and set all except the few with the highest SNR to zero;
\item Perform the second MCMC with all possible combinations of ``surviving'' wavelet coefficients to find the model with the least $\chi^2$ per degree of freedom (DoF)
\item Quantify the detection using $\Delta \chi^2/\Delta({\rm DoF})$, $\Delta$AIC and $\Delta$BIC.
\end{enumerate}

To start, we simulate future SN data for a dark energy model
with a feature in $w(z)$ given by
\be
w(z)= -1+A~{\rm exp}[-(z-z_{0})^2/\sigma^2] \ ,
\ee
where we choose $A=-0.9,~z_{0}=0.6,~
\sigma=0.1$. This $w(z)$ is shown with a black solid line in
Fig.~\ref{fig4}. We assume a flat universe, and choose the other
cosmological parameters as $\Omega_b = 0.046$, $\Omega_m = 0.279$
and $H_0 = 70 ~$km/s/Mpc. We also apply the distance priors derived from the CMB and BAO data as described in Sec.~\ref{sec:data}. We assume a SNAP-like JDEM survey with the
luminosity data binned into $17$ redshift bins in the range of
$z\in[0,1.7]$. Fig.~\ref{fig0} shows the comparison of the residuals (defined with respect to the $\Lambda$CDM model) of the distance modulus for the model with a bump-like feature in $w(z)$ and a constant-$w$ model which has the same average value $w$ (i.e. one that gives the same distance to the CMB last scattering surface). Error-bars are plotted according to the characteristics of SNAP.  The plot shows that a SNAP-like experiment is capable of differentiating between these models around $z = 0.5$.

Having produced our mock datasets,
we perform a MCMC fit for all the $16$ wavelet parameters plus $\Omega_b h^2$, $\Omega_m h^2$ and $h$,
assuming a flat universe and flat priors for all the wavelet parameters.
Fig.~\ref{fig2} shows the marginalized $1$-D likelihood distributions
for the $16$ wavelet parameters, with Table~\ref{tab:ston} listing
the corresponding mean values and the 68\% CL error bars. As one can
see, for many wavelet parameters the standard
deviation is much larger than the mean value, indicating that they
are not sensitive to the presence of a feature in $w$. Such poorly
constrained parameters are thus redundant, so we keep only the
ones with the highest SNR. In order for the D4 wavelets to be able to describe a constant $w$, one should keep $P_1$. In addition, we keep the $2$-$3$ coefficients with the highest SNR. In our example, we keep parameters $P1$, $P5$ and $P12$.

\begin{figure}[tbh]
\includegraphics[scale=0.7]{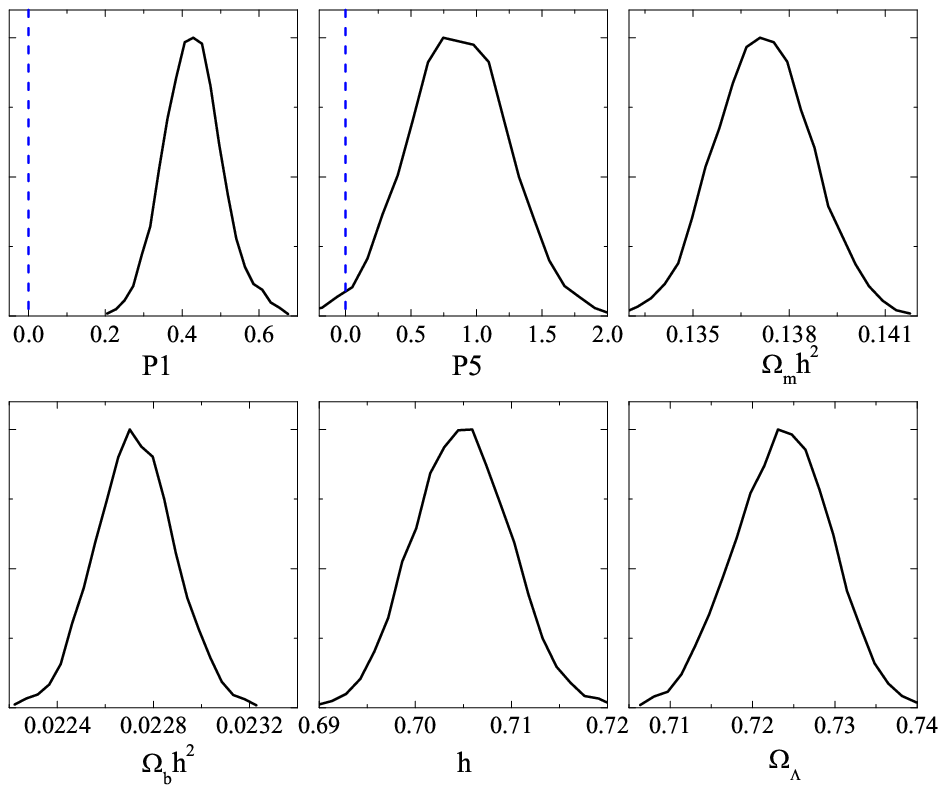}
\caption{$1$-D posterior distributions of the $1$st and the $5$th wavelet
coefficients, and the other four cosmological parameters.}
\label{fig3}
\end{figure}

\begin{figure}[tbh]
\includegraphics[scale=0.6]{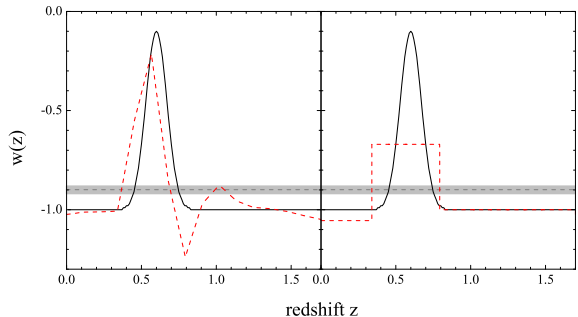}
\caption{The hidden feature (black solid), and the
reconstruction using D4 (red dashed, left panel) and Haar (red
dashed, right panel) wavelets. The horizontal dashed line and the
shaded band illustrate the fit with the constant $w$ model and the 68\%CL errors.}
\label{fig4}
\end{figure}

Next, we redo the MCMC using every combination of the selected
parameters (with $P1$ always included) and find the one that gives the minimum value of the
reduced $\chi^2$. In this case, the optimal combination is that of $P1$ and $P5$, while setting $P12=0$. Marginalized likelihood distributions for
this model are shown in Fig.~\ref{fig3} and the plot of $w(z)$
reconstructed using this model is shown in the left panel of
Fig.~\ref{fig4}.

\begin{table}[tbh]
\begin{tabular*}{0.45\textwidth}{c c c |c c c}
\hline \hline
 Para. & ~  Constraints~&  ~ SNR ~ & Para. & ~ Constraints ~ & ~ SNR\\
\hline
P1            & $-0.1\pm1.2$ & 0.1&   P9 & $-0.5\pm3.0$& 0.2 \\
P2            & $1.1\pm4.4$  & 0.3&   P10& $-0.7\pm4.3$& 0.2\\
P3            & $-1.3\pm4.3$ & 0.3&   P11& $1.5\pm4.6$ & 0.3\\
P4            & $-0.2\pm1.4$ & 0.1&   \underline{P12}& $\underline{7.4\pm9.2}$ & \underline{0.8}\\
\underline{P5}& $\underline{1.7\pm2.7}$  & \underline{0.6}&   P13& $-0.1\pm5.0$ & 0.0\\
P6            & $-0.3\pm4.7$ & 0.1&   P14& $0.1\pm5.2$ & 0.0\\
P7            & $0.2\pm5.2$  & 0.0&   P15& $1.6\pm4.8$ & 0.3\\
P8            & $-0.3\pm1.7$ & 0.2&   P16& $-1.6\pm4.1$ & 0.4\\
\hline \hline
\end{tabular*}
\caption{Marginalized mean values with 68\% CL. errors and SNR for
all the 16 wavelet parameters. The high SNR parameters are
underlined.} \label{tab:ston}
\end{table}

\begin{table}[tbh]
\begin{tabular} {cccc}
\hline
\hline
Model&$\Delta\chi^2/\Delta({\rm DoF})$ & $\Delta$AIC & $\Delta$BIC \\
\hline
 WCDM($w = -0.9$)  &0       &0      &0 \\
 CDM-D4-P1,5       &-8.8    &-6.3   &-3.7\\
 CDM-Haar-P2,4     &-8.3    &-6.8   &-3.2\\
\hline\hline
\end{tabular}
\caption{The differences in the model selection criteria for the simulated data with a bump feature in $w(z)$.}
\label{tab:model-sel-future}
\end{table}

In the above example two wavelet parameters were measured to be
non-zero. Their
uncorrelated linear combinations, or principal components, have
essentially the same values:
\be
PC1 = 0.9 \pm 0.4 ~~,~~ PC2 = 0.43 \pm 0.08 \ ,
\ee
indicating evidence for $\sim$6$\sigma$ and
$\sim$2$\sigma$ departures from zero respectively. The correlation
between the two wavelet parameters is negligible, which need not happen in general, but is somewhat expected.
Namely, this combination is the one most favored by the reduced $\chi^2$, which means that both parameters are likely needed to describe the data -- one of them cannot compensate for the other. This implies a
small correlation between these two parameters.

Detecting a departure from $w=-1$, which is all we have discussed so far, does not in itself amount to detecting
the bump feature. The value of $w(z)$ in Fig.~\ref{fig4} averaged
over redshift is less than $-1$, so we would detect a departure from
$-1$ by simply fitting a constant $w$. Hence, the more relevant question
is not to what extent wavelets are able to detect a departure from
$-1$, but to what extent the wavelet fit is preferred to a fit by a
simpler model with one constant $w$ parameter.

To answer this question, we fit the model with a constant $w$ (WCDM)
in a flat universe and apply the model selection criteria to determine the degree to which the wavelet model is preferred in comparison. In particular, we
calculate the change in $\chi^2$ per additional degree of freedom,
the difference in the BIC and AIC. As one see from the numbers in
Table~\ref{tab:model-sel-future}, the wavelet fit is strongly preferred according to all three criteria.

In addition to using D4 wavelets, we have followed the same
procedure and tried to recover the bump feature using the Haar
wavelets. We find again that a model with two wavelet parameters is
optimal and is preferred to WCDM at a similar significance level as
in the D4 case. The plot of $w(z)$ reconstructed using two Haar
wavelets is plotted in the right panel of Fig.~\ref{fig4}.

As a test, we have repeated the same procedure using data simulated
according to parameters of existing data sets. We confirmed that
the bump feature cannot be detected today at any significance.
We have also checked if we can accurately recover $w(z)$ if
$\Lambda$CDM was the true model for the universe. In this case all
the wavelet coefficients should be zero or the difference from zero
should not be statistically significant. To test that, we consider
simulated data for $\Lambda$CDM model and followed the same
procedure of selecting the first few coefficients with the highest
signal to noise, and then trying all different combinations of the
these selected wavelets. Our MCMC results in this case show no
evidence of detection of departure from zero for any of the wavelet
coefficients, and $\Lambda$CDM is correctly recovered as the best
model.

\begin{figure*}[tbh!]
\includegraphics[scale=0.8]{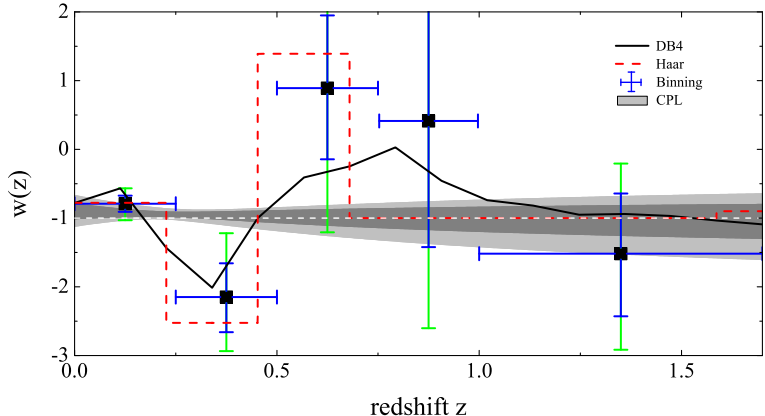}
\caption{Reconstruction of $w(z)$ from current data using D4 (black
solid line), Haar (red dashed line) wavelts, binning (error bars)
and CPL parametrization (shaded regions). The inner blue error bars
(inner shaded dark grey band) and outer green error bars (outer
shaded light grey band) show the 68 and 95\% CL. constraints
respectively.} \label{fig5}
\end{figure*}

\section{Searching for features in current data}
\label{sec:current}

In this section, we apply our method to the current SNe, CMB and BAO
data to search for the deviation of $w(z)$ from $-1$. We
use the recently released ``Constitution'' SNe data sample with 397
data points \cite{Hicken}, apply the distance priors measured by the
WMAP5 team \cite{wmap5-komatsu}, and assume a flat universe.

Table~\ref{tab:model-sel-current} shows the main results. We find
that using $16$ D4 or Haar wavelet coefficients reduces the $\chi^2$
and gives a hint that $w$ may have deviations from $\Lambda$CDM.
However, a model with $16$ parameters is not preferred by any of the
information criteria. With D4 wavelets, parameters $P1$, $P4$, and
$P8$ give the optimal model with a $\Delta \chi^2 / \Delta({\rm
DoF})$ of $-1.6$. With Haar wavelets, this number is $-1.9$ with the
three wavelet parameters being $P2$, $P4$ and $P5$. We also find
that none of the wavelet fits are favored by the AIC and BIC.

Plots of $w(z)$ reconstructed using the wavelet method with D4 and
Haar are shown in Fig.~\ref{fig5}. We also plot the results obtained
using the binning method and the CPL parametrization taken from
\cite{zhao}. As one can see, the best fit wavelet models show weak
hints of dark energy dynamics. In particular, $w(z)<-1$ for
$z<\in[0.25,0.5)$ and $w>-1$ for $z\in[0.5,1]$, implying that $w$
crosses $-1$ during its evolution. This is consistent with the
analysis of the ``Constitution'' data set in \cite{zhao}, which
reported weak evidence for a similar evolution history. Regardless
of whether such evolution is real or not, the CPL parametrization
would not be able to see it.

A significant deviation from
$0$ for any of the wavelet parameters shows deviation from
$\Lambda$CDM. Table~\ref{tab:pca} shows the eigen-values of
principal components for each model together with their errors,
separately. There is a $1 \sigma$ and $2 \sigma$
detection from D4 and Haar respectively, which can be interpreted as
weak hints that $w$ may have a departure from $-1$.

We note that although Ref.~\cite{zhao} obtained similar results, they
used a $5~w$ bin model. They found this optimal $5$-bin model after
performing a model-selection on $10$ different binning models, which is
computationally expensive. With the wavelet approach, we recover the same feature in
$w(z)$ with only $3$ wavelet parameters after running MCMC just a few times. This is possible because
of the multi-resolution nature of the wavelets. We should also note that systematic effects like Malmquist bias \cite{Kowalski:2008ez} can mimic the same behaviour as dynamical $w$. Therefore, we need to be cautious about making any strong conclusions about the equation of state parameter $w$.

\begin{table}[tbh]
\begin{tabular} {cccc}
\hline
\hline
Model&$\Delta\chi^2/\Delta({\rm DoF})$ & $\Delta$AIC & $\Delta$BIC \\
\hline
$\Lambda$CDM  & 0            &  0  & 0\\
CDM-D4-16Ps       & -0.3  & 32.7   & 91\\
CDM-Haar-16Ps        & -0.4      & 31.1 & 89\\
\hline
CDM-D4-1,4,8     &-1.6       & 0.9 & 13\\
CDM-Haar-2,4,5    &-1.9       & 0.2 & 12.3\\
\hline\hline
\end{tabular}
\caption{Differences in the model selection criteria for the analysis of the current data.}
\label{tab:model-sel-current}
\end{table}

\begin{table}[tbh]
\begin{tabular}{cccc}
\hline \hline
Model             & PC1              & PC2              & PC3 \\
\hline
D4-1,4,8          & 0.25$\pm$1.04    &  0.36$\pm$0.39   & 0.07$\pm$0.13 \\
Haar-2,4,5       &-4.03$\pm$1.81    & -0.26$\pm$0.46   & 0.10$\pm$0.15\\
\hline
\end{tabular}
\caption{Principal Components for wavelet coefficients of optimum
$w$ models with current cosmological data}
\label{tab:pca}
\end{table}

\section{Summary}
\label{sec:summary}

We investigated the utility of the wavelet expansion method for
recovering a local redshift-dependence in the dark energy equation of state
$w(z)$ with future cosmological data. The advantage of
wavelets over other methods is that there is no need to assume a
scale or a position for a feature in $w(z)$. With wavelets, one is able to search for small scale features, while looking at the large scale picture at the
same time. Our results show that a bump-like feature in $w(z)$, which is not ruled out by today's data,
can be recovered using
wavelets at a high significance with future data.

We have tried relaxing the flatness assumption and considered curvature as an additional parameter. In this case, we find
that the bump feature we put in becomes undetectable, even with
the wavelet technique, due to the degeneracy between $w$ and the
curvature $\Omega_K$. There is sufficient freedom for
$\Omega_K$ and $\Omega_m$ to adjust themselves to mimic the effect of
a featured $w(z)$ in the luminosity distance observations.
More specifically, we find that a model with $ w = -1.2$ ,
$\Omega_m h^2 = 0.137$, $\Omega_b h^2 = 0.0227$, $\Omega_\Lambda = 0.6$, $\Omega_K = 0.1$
and $h = 0.68$ has nearly the same goodness-of-fit as the
bump model.

Applying the wavelet method to current cosmological data
we find a hint of departure from $\Lambda$CDM at a 2$\sigma$ level.
The same behavior has been found independently using the binning
method \cite{zhao} at the same significance level. We have shown
that such local features can be found more efficiently by using wavelets.
The feature we found in current data will require further investigation as the SNe and
other cosmological data improves.

\acknowledgments
This work was supported by an NSERC Discovery Grant and funds from SFU.


\begin{thebibliography}{99}

\bibitem{Riess:1998cb}
  A.~G.~Riess {\it et al.}  [Supernova Search Team Collaboration],
  Astron.\ J.\  {\bf 116}, 1009 (1998)
  [arXiv:astro-ph/9805201].

\bibitem{Perlmutter:1998np}
  S.~Perlmutter {\it et al.}  [Supernova Cosmology Project Collaboration],
  Astrophys.\ J.\  {\bf 517}, 565 (1999)
  [arXiv:astro-ph/9812133].

\bibitem{wmap5-komatsu}
  E.~Komatsu {\it et al.}  [WMAP Collaboration],
  Astrophys.\ J.\ Suppl.\  {\bf 180}, 330 (2009)
  [arXiv:0803.0547 [astro-ph]].

\bibitem{albrecht_detf} 
  A.~J.~Albrecht {\it et al.},
  arXiv:astro-ph/0609591.

\bibitem{Linder:2002et}
  E.~V.~Linder,
  Phys.\ Rev.\ Lett.\  {\bf 90}, 091301 (2003)
  [arXiv:astro-ph/0208512].

\bibitem{Chevallier:2000qy}
  M.~Chevallier and D.~Polarski,
  Int.\ J.\ Mod.\ Phys.\  D {\bf 10}, 213 (2001)
  [arXiv:gr-qc/0009008].

\bibitem{corasaniti}
  P.~S.~Corasaniti, M.~Kunz, D.~Parkinson, E.~J.~Copeland and B.~A.~Bassett,
  Phys.\ Rev.\  D {\bf 70}, 083006 (2004)
  [arXiv:astro-ph/0406608].

\bibitem{alam}
  U.~Alam, V.~Sahni, T.~D.~Saini and A.~A.~Starobinsky,
  Mon.\ Not.\ Roy.\ Astron.\ Soc.\  {\bf 354}, 275 (2004)
  [arXiv:astro-ph/0311364].

\bibitem{pca-huterer}
  D.~Huterer and G.~Starkman,
  Phys.\ Rev.\ Lett.\  {\bf 90}, 031301 (2003).

\bibitem{pca-pogosian}
  R.~G.~Crittenden, L.~Pogosian and G.~Zhao,
  JCAP \ {\bf 12}, 025 (2009)
  [arXiv:astro-ph/0510293]

\bibitem{pca-cooray}
  D.~Huterer and A.~Cooray,
  Phys.\ Rev.\  D {\bf 71}, 023506 (2005)
  [arXiv:astro-ph/0404062].

\bibitem{Starobinsky}
  T.~D.~Saini, S.~Raychaudhury, V.~Sahni and A.~A.~Starobinsky,
  Phys.\ Rev.\ Lett.\  {\bf 85}, 1162 (2000)
  [arXiv:astro-ph/9910231].

\bibitem{jdem_fom} 
  A.~J.~Albrecht {\it et al.},
  arXiv:0901.0721 [astro-ph.IM].

\bibitem{Carroll:2003st}
  S.~M.~Carroll, M.~Hoffman and M.~Trodden,
  Phys.\ Rev.\  D {\bf 68}, 023509 (2003)
  [arXiv:astro-ph/0301273].

\bibitem{Mukherjee}
  P.~Mukherjee and Y.~Wang,
  Astrophys.\ J.\  {\bf 598}, 779 (2003)
  [arXiv:astro-ph/0301562].

\bibitem{Shafieloo2}
  A.~Shafieloo, T.~Souradeep, P.~Manimaran, P.~K.~Panigrahi and R.~Rangarajan,
  Phys.\ Rev.\  D {\bf 75}, 123502 (2007)
  [arXiv:astro-ph/0611352].

\bibitem{wavelet_nonG}
  M.~Hobson, A.~Jones and A.~Lasenby,
  arXiv:astro-ph/9810200;
   P.~Mukherjee, M.~P.~Hobson and A.~N.~Lasenby,
  Mon.\ Not.\ Roy.\ Astron.\ Soc.\  {\bf 318}, 1157 (2000)
  [arXiv:astro-ph/0001385];
    P.~Mukherjee and Y.~Wang,
  Astrophys.\ J.\  {\bf 613}, 51 (2004)
  [arXiv:astro-ph/0402602];
   J.~D.~McEwen, M.~P.~Hobson, A.~N.~Lasenby and D.~J.~Mortlock,
  Mon.\ Not.\ Roy.\ Astron.\ Soc.\ Lett.\  {\bf 371}, L50 (2006)
  [arXiv:astro-ph/0604305];
    A.~Curto, E.~Martinez-Gonzalez, P.~Mukherjee, R.~B.~Barreiro, F.~K.~Hansen, M.~Liguori and S.~Matarrese,
  arXiv:0807.0231 [astro-ph];
  P.~Cabella {\it et al.},
  arXiv:0910.4362 [astro-ph.CO].


\bibitem{Cabella:2007br}
  P.~Cabella, P.~Natoli and J.~Silk,
  Phys.\ Rev.\  D {\bf 76}, 123014 (2007)
  [arXiv:0705.0810 [astro-ph]].



\bibitem{Daubechies}
Daubechies, I. 1988, Communications on Pure and Applied Mathematics, vol. 41, pp. 909996

\bibitem{NR}
Numerical Recipes 3rd Edition: The Art of Scientific Computing by William H. Press, Saul A. Teukolsky, William T. Vetterling, and Brian P. Flannery

\bibitem{AIC}
Akaike, Hirotugu (1974). "A new look at the statistical model identification". IEEE Transactions on Automatic Control 19 (6): 716723. 

\bibitem{BIC}
Schwarz G., 1978, Ann. Statist., {\bf 5}, 461

\bibitem{kim_etal}
  A.~G.~Kim, E.~V.~Linder, R.~Miquel and N.~Mostek,
  Mon.\ Not.\ Roy.\ Astron.\ Soc.\  {\bf 347}, 909 (2004)
  [arXiv:astro-ph/0304509].

\bibitem{nsnf} G.~Aldering {\it et al}, Proc. SPIE, 4836, 61 (2002).

\bibitem{Hicken}
  M.~Hicken {\it et al.},
  Astrophys.\ J.\  {\bf 700}, 1097 (2009)
  [arXiv:0901.4804 [astro-ph.CO]].


\bibitem{wang}
  Y.~Wang and P.~Mukherjee,
  Phys.\ Rev.\  D {\bf 76}, 103533 (2007)
  [arXiv:astro-ph/0703780].



\bibitem{Elgaroy}

  O.~Elgaroy and T.~Multamaki,
  arXiv:astro-ph/0702343.

\bibitem{Li:2008cj}
  H.~Li, J.~Q.~Xia, G.~B.~Zhao, Z.~H.~Fan and X.~Zhang,
  Astrophys.\ J.\  {\bf 683}, L1 (2008)
  [arXiv:0805.1118 [astro-ph]].


\bibitem{planck}

  P.~Mukherjee, M.~Kunz, D.~Parkinson and Y.~Wang,
  Phys.\ Rev.\  D {\bf 78}, 083529 (2008)
  [arXiv:0803.1616 [astro-ph]].


\bibitem{Percival}

  W.~J.~Percival, S.~Cole, D.~J.~Eisenstein, R.~C.~Nichol, J.~A.~Peacock, A.~C.~Pope and A.~S.~Szalay,
  Mon.\ Not.\ Roy.\ Astron.\ Soc.\  {\bf 381}, 1053 (2007)
  [arXiv:0705.3323 [astro-ph]].

\bibitem{PCA}
Pearson, K. (1901). "On Lines and Planes of Closest Fit to Systems of Points in Space" (PDF). Philosophical Magazine 2 (6): 559572. 



\bibitem{bump}
  J.~Q.~Xia, G.~B.~Zhao, H.~Li, B.~Feng and X.~Zhang,
  Phys.\ Rev.\  D {\bf 74}, 083521 (2006).

\bibitem{zhao}
  G.~B.~Zhao and X.~Zhang,
  arXiv:0908.1568 [astro-ph.CO].


\bibitem{Kowalski:2008ez}
  M.~Kowalski {\it et al.}  [Supernova Cosmology Project Collaboration],
  Astrophys.\ J.\  {\bf 686}, 749 (2008)
  [arXiv:0804.4142 [astro-ph]].


\end{thebibliography}
\end{document}